\documentclass[letterpaper,english,preprint]{IEEEtran}
\usepackage[T1]{fontenc}
\usepackage[latin9]{inputenc}
\usepackage{graphicx}
\usepackage[numbers]{natbib}

\makeatletter

\providecommand{\tabularnewline}{\\}

\@ifundefined{showcaptionsetup}{}{%
 \PassOptionsToPackage{caption=false}{subfig}}
\usepackage{subfig}
\makeatother

\usepackage{babel}
\begin{document}

\title{NSense: A People-centric, non-intrusive Opportunistic Sensing Tool
for Contextualizing Nearness }

\author{Rute Sofia\thanks{Corresponding Author. COPELABS, University Lusofona Campus, Building
U, First Floor, Campo Grande 388, 1749-024 Lisboa.}, Saeik Firdose, Luis A. Lopes, Waldir Moreira, Paulo Mendes \\
\IEEEauthorblockA{COPELABS, University Lusofona\\
\{rute.sofia, saeik.firdose,luis.amaral,waldir.moreira,paulo.mendes\}@ulusofona.pt}}
\maketitle
\begin{abstract}
In the context of social well-being and context awareness several
eHealth applications have been focused on tracking activities, such
as sleep or specific fitness habits, with the purpose of promoting
physical well-being with increasing success. Sensing technology can,
however, be applied to improve social well-being, in addition to physical
well-being. This paper addresses NSense, a tool that has been developed
to capture and to infer social interaction patterns aiming to assist
in the promotion of social well-being. Experiments carried out under
realistic settings validate the NSense performance in terms of its
capability to infer social interaction context based on our proposed
computational utility functions. Traces obtained during the experiments
are available via the CRAWDAD international trace repository.
\end{abstract}


\section{Introduction}

As technology becomes more pervasive, low-cost sensing devices such
as smartphones are carried around by a large number of people, thus
giving rise to the opportunity of capturing diverse aspects of the
human routine and to take advantage of inferred patterns to improve
several social and behavioral aspects, thus promoting better living/well-being.
For instance, today there are several tools that bring awareness to
our fitness routine, or even to our daily activity patterns, e.g.,
by looking into sleep, motion, as well as other activity and behavioral
patterns. There are also several frameworks that attempt to gather
healthcare data in the cloud so as to assist in a better prevention
of diseases, or in a better control of conditions, such as asthma. 

Derived from advances in communication and in particular, in networked
systems, today the available technology can be relied upon to take
advantage of sensing as well as of direct communication via short-range
wireless technology (e.g. Bluetooth; Wi-Fi Direct) to assist in better
classifying our daily routines, in a way that is not necessarily intrusive,
and in a way that does not endanger one\textquoteright s personal
sphere. Such advantages are relevant in the context of promoting well-being
in its multiple dimensions (social, cognitive/mental, and physical),
and improving our daily routines (e.g. by stimulating social cohesion). 

NSense (\emph{Nearness Sense}) has been devised as a software-based
solution to promote social well-being (and consequently, to assist
in the development of frameworks and strategies that may provide benefits
to mental and to cognitive well-being), by exploring individual and
collective \emph{nearness} contextualization. In the category of non-intrusive
sensing, NSense is one of the first middleware solutions that exploits
the notion of nearness via the use of short-range wireless technology
(Wi-Fi Direct, Bluetooth) together with other sensors while integrating
an efficient sampling and storage strategy (e.g., not recurring always
to geo-positioning information; reducing microphone sampling), thus
achieving a sound balance in terms of storage and battery drainage
while performing behavior inference locally in the personal device,
without relying to external computational devices (e.g., cloud). Moreover,
aiming to ensure a large scale deployment, NSense is designed based
on a modular software architecture, which allows an easy integration
of other sensors.

This paper is structured as follows. Section \ref{sec:Related-Work}
goes over related work explaining how NSense introduces benefits in
the context of social well-being. Section \ref{sec:Inferring-a-Social}
discusses our perspective on contextualizing nearness, proposing as
valid markers for such context two utility functions: \emph{social
interaction}, that infers the interaction of devices over time; \emph{propinquity},
that measures the probability of social interaction occurring over
time, i.e., the probability of social ties to grow stronger over time
and space. The NSense software architecture is briefly presented in
Section \ref{sec:USense-Node-Architecture} which covers the computational
and specification aspects of this tool. Section \ref{sec:Experimentation-Analysis}
is dedicated to an evaluation of NSense in terms of its capability
to capture nearness under realistic conditions. Section \ref{sec:Conclusions}
concludes the paper with a summary of findings as well as ongoing
research directions.

\section{Related Work\label{sec:Related-Work}}

There are different proposals aiming at exploiting sensing data produced
by smartphones to infer individual behavior. Such proposals can be
divided into two major families\emph{, }namely, \emph{participatory}
sensing \citep{Lane08:urban}, where the user actively engages in
the data collection activity; \emph{opportunistic }sensing \citep{Lane10:survey},
where the data collection stage is fully automated with no user involvement,
i.e., in a non-intrusive fashion. While participatory and opportunistic
sensing have complementary properties, we believe that a pure opportunistic
sensing approach is more suitable to support large scale deployments
and application diversity. 

Most of the available sensing solutions are focused on the inference
of individual behavior. CenceMe \citep{CenceMe} has been designed
to allow users to share their sensing experience with trusted circles
on social networks. Such experience relates with activities such as
running; walking; disposition; contextual surroundings. EmotionSense
\citep{emotionsense} aims at correlating how the user feels with
user activities, and BeWell \citep{bewell1} aims at tracking the
impact of sleep, physical activity, and social interaction on the
user\textquoteright s well-being, where social interaction is based
on phone usage (SMS and calls). Both CenceMe and BeWell fall into
the category of non-intrusive sensing. BeWell integrates speech activity
detection to provide a better contextualization of the user's social
context. Nevertheless, the inference of conversational activities
done by BeWell is not enough to reliably assess a user social context
in terms of how socially engaged the owner of the device is with different
individuals in a given setting (i.e., home, school, work). 

Sociometer \citep{thesociometer} focuses on the notion of \emph{social
engagement} given by proximity and conversational activities to understand
how users interact. Sociometer is a mark in terms of better understanding
cues concerning social context and the fact that activity recognition
can be correlated with social engagement. Nevertheless, Sociometer
falls into the category of intrusive tools, where social interaction
is derived from infra-red proximity technology, which requires clear
line-of-sight, while social engagement can still take place with obstacle
between users and/or without users being facing one another. 

Similarly to Sociometer, SociableSense \citep{sociablesense} aims
at inferring individual behavior in the context of office environments.
SocialSense is based on a smartphone platform, thus being less intrusive.
SociableSense adaptively controls the sampling rate of accelerometer,
Bluetooth, and microphone sensors in order to estimate the user's
sociability, and strength of relationship with colleagues. 

With NSense, we aim at inferring nearness based on the user's daily
routine and social interaction over time and space.

\section{Inferring Nearness\label{sec:Inferring-a-Social}}

\subsection{Terminology and Notation}

\label{sec:terminology}

This section provides terminology that is required for a better understanding
of the paper. NSense is middleware that has been designed to run on
a personal sensing device such as a smartphone. As such, a \emph{user}
corresponds to the device \emph{owner} and \emph{carrier}. A device
is owned by a person only, and the identification of the device is
associated to its owner only. 

\emph{Propinquity} \citep{Reagans2011} is used in this work as an
indicator of the probability of devices (and hence, their carriers)
to strengthen over time their nearness. Propinquity refers to the
physical and/or psychological proximity between people. It has been
studied mostly to understand how interpersonal relations develop within
the same space. For instance, people living in the same floor of a
building attain a higher propinquity than those living on different
floors. Two people sharing similar beliefs also attain a higher propinquity
than those that do not share beliefs. Hence, propinquity is a property
that is highly relevant to consider when defining, in networked systems,
social interaction contextualization. Owners of devices that are in
close range to each other, or that meet often, are expected to have
a higher propinquity than owners of devices that do not meet often.
Other aspects, such as environmental sound, or distance between devices,
can assist in inferring propinquity.

\emph{Social interaction} provides an indication on how much owners
of devices have been interacting over time and space, derived from
aspects such as the distance between such devices; the sound level
activity around the devices; the type of movement of the devices. 

Propinquity and social interaction assist in tracking nearness derived
from the natural networking footprint that devices and their human
carriers leave around. In other words, these two utility functions
are relevant to develop solutions that capture nearness in a way that
is non-intrusive and that does not jeopardize in any way the personal
sphere of citizens.

The parameters considered by NSense in the definition of a nearness
context concern \emph{node degree}; \emph{node motion}; \emph{social
strength}; relative distance; and ambient sound level. The node degree
$n(i)_{t}$ at an instant $t$ is commonly used to characterize several
aspects of networked systems. A deterministic variable, $m(i)_{t}$
, provides an indication is the node is moving based on the motion
on a three accelerometer axis. The social\emph{ strength }of node
$i$ towards node $j$\emph{ }in a specific hourly sample $h$ , for
day $d$\emph{, $s(i,j)_{d,h}$} \citep{Moreira2012g} is derived
from contact duration between nodes $i$ and $j$ during a specific
time window $h$ in a passive way. The relative distance between devices
$i$ and $j$, $d(i,j)_{t}$, corresponds to an exponential moving
average of the euclidean distance between the two nodes, following
a propagation loss model. Finally, the environmental sound $nl(i)_{t}$,
measured by the device based on a sound activity detection algorithm,
is used to classify the environmental sound context based on noise
levels with the help of the deterministic variable $v$. 

\subsection{Relating Classified Activities with Nearness}

This section explains how the different parameters are sensed by NSense
and how we believe that such sensing can assist in inferring, with
a reasonable level of accuracy, a nearness context, eventually leading
to ways to stimulate nearness, if intended. A more detailed explanation
of NSense, its pipelines, as well as aspects concerning classification
of activities and our perspective concerning relation to nearness
is available via a more detailed technical report \citep{R.C.SofiaS.FirdoseL.A.LopesW.Moreira2016}.

Nearness considers both psychological (social) proximity as well as
physical proximity aspects. It is here assumed that the level of social
proximity derived from portable devices carries a correspondence to
the level of social interaction that devices can capture via sensing
interfaces. It is also assumed that the classification of aspects
related to individual behavior, such as distance towards neighboring
devices, motion, as well as environmental sound level assists in a
more accurate classification of nearness aspects.

A nearness context is therefore modeled by relying on the two utility
functions explained in the previous section. While the social interaction
utility function provides a realistic measure on how social interaction
is occurring, based on physical proximity metrics, the propinquity
function operates as a reinforcement that social interaction grows
with time - stronger times are formed when propinquity is higher,
over time and space \citep{Reagans2011}.

For the purpose of relying on non-intrusive opportunistic sensing,
it is assumed that devices hold the proposed set of sensors, and that
their owners allow data capture in accordance with the aspects that
have been described in the previous sections. Table \ref{tab:Potential-cases-and}
illustrates a potential correlation between the different activities
sensed, and how such sensing can impact on the modeled propinquity
as well as on the modeled social interaction. 

\begin{table}
\caption{Empirical cases and results for correlation between classification
and inference.\label{tab:Potential-cases-and}}

\centering{}{\scriptsize{}}%
\begin{tabular}{|c|c|c|c|c|}
\hline 
{\scriptsize{}Activity} & {\scriptsize{}Parameter} & {\scriptsize{}1} & {\scriptsize{}2} & {\scriptsize{}3}\tabularnewline
\hline 
\hline 
{\scriptsize{}Social strength (Proximity)} & {\scriptsize{}$s(i,j)$} & {\scriptsize{}High} & {\scriptsize{}High} & {\scriptsize{}Low}\tabularnewline
\hline 
{\scriptsize{}Sound activity detection} & {\scriptsize{}$v(i)$} & {\scriptsize{}Quiet} & {\scriptsize{}Alert} & {\scriptsize{}Quiet}\tabularnewline
\hline 
{\scriptsize{}Relative Distance} & {\scriptsize{}$d(i,j)$} & {\scriptsize{}Short} & {\scriptsize{}Large} & {\scriptsize{}Short}\tabularnewline
\hline 
{\scriptsize{}Motion} & {\scriptsize{}$m(i)$} & {\scriptsize{}Stationary} & {\scriptsize{}Stationary} & {\scriptsize{}Moving}\tabularnewline
\hline 
\textbf{\scriptsize{}Social Interaction } & \textbf{\scriptsize{}$si(i,j)$} & {\scriptsize{}Avg} & {\scriptsize{}Low} & {\scriptsize{}Low}\tabularnewline
\hline 
\textbf{\scriptsize{}Propinquity} & \textbf{\scriptsize{}$p(i,j)$} & {\scriptsize{}High} & {\scriptsize{}High} & {\scriptsize{}High}\tabularnewline
\hline 
{\scriptsize{}Nearness} & {\scriptsize{}-} & {\scriptsize{}Avg} & {\scriptsize{}Avg} & {\scriptsize{}Avg}\tabularnewline
\hline 
\end{tabular}{\scriptsize\par}
\end{table}

For the sake of explanation, in the provided examples we assume that
the social strength has either high or low values, while the level
of sound is determined by either quiet or alert values. Likewise,
the relative distance is considered to be either short or large. In
terms of motion nodes are classified as being stationary or moving. 

On a first case (cf. Table \ref{tab:Potential-cases-and}, (1)) two
nodes $i$ and $j$ meet frequently and hence their social strength
is high. They are close by (short distance, e.g., 10 meters) as well
as stationary. The sound activity level surrounding them is low (\emph{quiet}).
This seems to imply that there are conditions to strengthen nearness
for the people carrying the devices. For instance, this could occur
if two strangers share frequently the same bus ride. While if instead
social interaction would be high, then it could be the case of two
friends quietly reading at the same table on a coffee-shop. It should
be highlighted that currently NSense is simply detecting environmental
sound levels only.

In case 2 it is assumed that the distance changes, thus impacting
both propinquity as well as social interaction. Nearness remains average,
as propinquity has a high value.

In terms of the impact of motion in the inferrence of a nearness context
(case 3), we assume that when distances between nodes are kept short,
independently of whether moving or not, propinquity is high. For this
specific example, social interaction is low as the social strength
between the devices is also low. Nearness between the people can nevertheless
still be classified as average. 

These simplistic examples aim at explaining why a nearness context
requires considering both social interaction as well as propinquity
heuristics. For the next sections, to quantify propinquity we consider
Eq. \ref{eq:propinquity}.

{\footnotesize{}
\begin{equation}
p(i)=s(i,j)_{t}*\frac{1}{(d(i,j)_{t}+1)*m(i)_{t}}\label{eq:propinquity}
\end{equation}
}{\footnotesize\par}

Propinquity has been modelled to be directly proportional to the social
strength, and inversely proportional to the distance and motion of
the node. Based on our analysis as well as on prior work \citep{Reagans2011},
the current sound level activity is not relevant to model propinquity,
as this function assists in understanding how social interaction can
become stronger (stronger ties). We have then modeled social interaction
as in Eq. \ref{eq:social interaction}, where{\footnotesize{}$\:$$\sigma^{2}=0.75$
}is the standard deviation for a normal distribution based on{\footnotesize{}$\:$}$v$,
and{\footnotesize{}$\:$}$\mu${\footnotesize{}$\:$} is the mean
for such distribution{\footnotesize{}.}{\footnotesize\par}

{\tiny{}
\begin{equation}
si(i,j)_{t}=\log(s(i,j)_{t})*\frac{1}{\sigma*\sqrt{2\Pi}}*e^{\frac{-(v-\mu)^{2}}{2*\sigma^{2}}}*\frac{1}{log(d(i,j)_{t}+10)*m(i)_{t}}\label{eq:social interaction}
\end{equation}
}{\tiny\par}

\section{NSense Architecture\label{sec:USense-Node-Architecture}}

\begin{figure}
\begin{centering}
\includegraphics[scale=0.35]{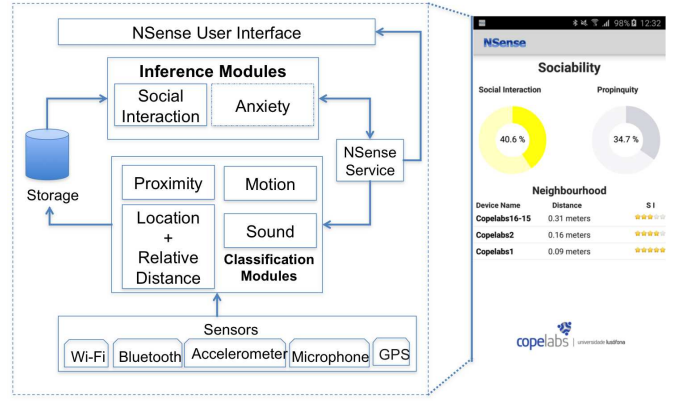}
\par\end{centering}
\centering{}\caption{\label{fig:USense-node-architecture.-1}NSense node architecture.}
\end{figure}

The high-level architecture of NSense is illustrated in Figure \ref{fig:USense-node-architecture.-1},
where dashed squares represent modules that are not yet implemented
(self-reporting and inference of anxiety levels) or that have been
implemented but are not considered in this paper (roaming behavior
classification \citep{Sofia2015}). 

NSense is an open-source tool licensed under LGPLv3.0 \citep{usense2015}
and has been designed to rely on data captured via multiple sensors
(Wi-Fi, Bluetooth, accelerometer, microphone), data which is then
worked via utility functions that classify activity in terms of social
proximity; relative distance; location; motion; as well as surrounding
sound level. 

Activity classification is then used to infer about the behavior of
the person carrying/owning a device. For the purpose of this paper
the NSense implementation and experiments have been focused on social
interaction contextualization only. 

The \emph{NSense Service} is the component responsible for initiating
all of the software modules as well as all configured sensors, ensuring
the modularity necessary for the easy plug-in of future sensors and
classification modules. The NSense service is also responsible for
controlling data access in the local SQLite database, as well as,
if required, to dump data in a remote database. 

The values required for the inference of behavior are computed by
independent classification modules named \emph{pipelines}, where a
pipeline corresponds to a set of operations performed by a classification
module done over a set of sensors. Classification modules that use
more than one sensor, as is the case of the location module, and sensors
that are used by several modules, as is the case of Wi-Fi. 

Each pipeline captures raw data from a set of sensors and, without
storing it, transforms such data via the application of specific utility
functions into people-centric actions: ranking of preferred locations;
level of surrounding noise; relative distance towards current neighboring
devices, social strength towards any encountered devices; motion status
of the device. Only the result of the pipeline computation is kept
on the database, thus reducing storage and energy consumption required
in opportunistic sensing.

The operation of NSense is based on four pipelines, each one encompassing
sensing and classification activities. The location pipeline, based
on the MTracker tool \citep{Sofia2015}, performs non-intrusive classification
of the location of the device. This classification is based on the
visited wireless networks, combined with information collected from
the network operator and GPS, as well as the relative distance between
neighboring devices based on Wi-Fi direct.

The proximity pipeline relies on data captured via Bluetooth to estimate
the social strength of a device towards all other encountered devices.
This classification is based on the duration of each encounter. The
social strength is the NSense variable that provides information about
the user's hourly routine through different days. Although the proximity
pipeline considers only the Bluetooth interface, the next version
of NSense will also consider data captured via Wi-Fi direct.

The motion and sound activity detection pipelines follow an \emph{Activity
Recognition Chain (ARC)} model, which comprises stages for data acquisition,
signal pre-processing and segmentation, feature extraction and selection,
training, and classification. The motion pipeline relies on data captured
via the accelerometer to perform a classification of the type of movement
activity in terms of being stationary or moving, capturing the percentage
of time that a device is either stationary or moving. The sound activity
detection pipeline captures the level of environmental sound derived
from the sound amplitude, to classify in a deterministic way the environmental
sound level. 

\section{Performance Analysis\label{sec:Experimentation-Analysis}}

This section covers experimentation that we have done with NSense
in order to evaluate a nearness context based on the proposed propinquity
and social interaction functions, over time and space. The purpose
of this performance analysis is to show that the functions are robust
enough to characterize nearness under realistic settings. The experiments
consider 4 Samsung S3 devices running Android 4.2, having NSense installed
as background service. Over different days and time periods, the devices
have been carried by users that share affiliation. The devices are
identified as USense2, USense3, USense4, and USense5\footnote{The designation of the devices used in the experiment reflect the
nomenclature that NSense had at the time of the experiments, i.e.,
USense.}. The devices continuously sensed data which has been recorded every
minute for different time periods. For instance, during the first
experiment we have considered a 7-hour period, while on the second
experiment we have considered a 50-hour period. On the last experiment
we have considered a smaller subset of traces extracted over 3 hours.
The experiments have been repeated several times over different days,
for 3 weeks.\footnote{Traces are available via CRAWDAD, http://crawdad.org/copelabs/usense/,
or http://copelabs.ulusofona.pt/scicommons/index.php/publications/show/844.}

On a first set of experiments (\emph{Experiment I}) we have analyzed
the capability and robustness of the functions proposed to capture
the correlation of the different sensed activities over time and space,
for a short period of time (7 hours, from 8 a.m. to 3 p.m. GMT). Then,
on a second set of experiments (\emph{Experiment II}) we have increased
such period to 50 hours to understand whether or not the functions
would still be robust. In both sets of experiments the results are
presented from the perspective of a specific node, even though data
has been obtained from the perspective of all nodes. On a third set
of experiments (\emph{Experiment III}) we have analyzed properties
of social proximity and propinquity derived from experiments between
2 nodes, for a period of 3 hours.

\subsection{Experiment I: Nearness Inference Context, Short Time Period}

\begin{figure}
\center
\vspace{-1cm}\includegraphics[scale=0.4]{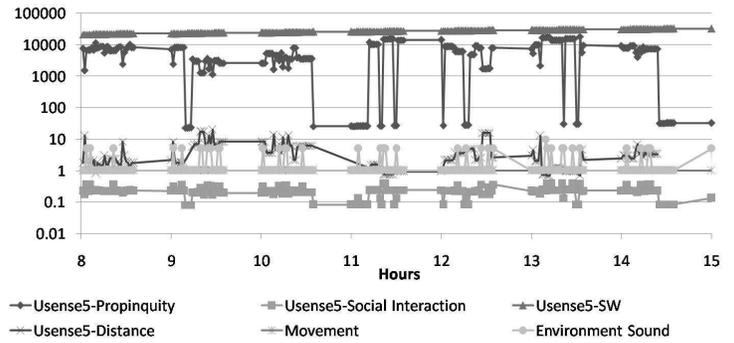}\\
\caption{Experiment I, USense2 towards USense5.\label{fig:Experiment-I-results.}}
\end{figure}

On a first scenario we have considered the perspective of USense2
device in regards to the three other devices over a period of 22 hours,
of which we have extracted a period of 7 hours, from 8a.m. to 3p.m,
having samples been obtained every minute. Results presented in Figure
\ref{fig:Experiment-I-results.} concern the perspective of device
USense2 towards USense5 \footnote{\tiny{Due to space constrains, we have selected representative examples
of all the experiments carried out. The full set of results is available
via \citep{R.C.SofiaS.FirdoseL.A.LopesW.Moreira2016}}.}. The carriers of these devices share affiliation. For each figure,
the X-axis represents the hours, while the Y-axis provides the results
attained in a logarithmic scale. 

Propinquity and social interaction remain stable, while the social
weight slightly increases over time. For the instants when distance
increases, propinquity lowers, as the probability of having nodes
around is lower. A downgrade in social interaction is also observable,
even though such downgrade is less significant than in propinquity,
when both sound levels go down, and distance increases, as this means
that the device may be getting isolated. Sound impacts social interaction
more than propinquity, as sound is relevant to understand whether
or not the owners of close-by devices are truly interacting.

The social weight between the devices varies slightly over the period
observed. The social weight measures the interaction over time and
takes into consideration the fact that the devices may have frequently
met in the past. While propinquity and social interaction consider
the current instant. 

\subsection{Experiment II: Nearness Inference Context, Long Time Period}

\begin{figure}
\center
\vspace{-1cm}\includegraphics[scale=0.35]{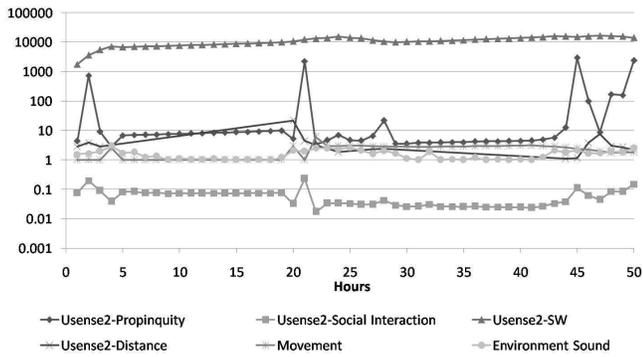}

\caption{Experiment II results, USense5 towards USense2.\label{fig:Experiment-II-results}}
\end{figure}

Experiment II concerns a longer period of observation, based on another
source node, USense5. Data has been captured over a 50 hour period,
and while on the prior set of experiments the source node (USense2)
was stationary most of the time, in this experiment USense5 has been
carried around. Samples have been collected again every minute between
23.11.2015, 2 p.m. GMT, and 25.11.2015, 4 p.m. GMT; however, for achieving
a better visualization, the results shown in the charts of Figure
\ref{fig:Experiment-II-results} are shown per hour. For computing
distance over time, the readings that could not be obtained were discarded.The
X-axis corresponds to hours, where hour 0 corresponds to 23.11.2015,
2 p.m., and hour 50 corresponds to 25.11.2015, 4 p.m. The results
in the Y-axis are presented in logarithmic scale. 

The variations in social interaction and propinquity as well as in
social weight tell us that the two devices have been interacting on
the first two observed hours (period between 23.11.2015, 2 p.m. and
4 p.m.). Between hours 5 and 20 the device has been stationary and
distance readings towards Usense2 were not available. This means that
the two devices were not interacting during that period as captured
by the two proposed functions (propinquity and social interaction).
Then, at hour 21 propinquity increases, due to the fact that the devices
met again and are at a quite close range. Between hours 20 and 30
(24.11.2015, 10 a.m. and 24.11.2015, 8 p.m.) the devices interact.
However, the environmental sound seems to be low. This implies that
nearness is high (and hence, propinquity varies), and yet, social
interaction may be low (devices are in close range and yet, environmental
sound is low). During the night period the interaction varies little
as expected, resuming around hour 45 (around 25.11.2015, 10 a.m.). 

\subsection{Experiment III: Asymmetry in a Nearness Context}

One aspect that we want to understand is whether or not the social
interaction and propinquity patterns can be correlated based on individual
neighbor\textquoteright s perspective as well as based on a global
perspective of nodes interacting. In other words: if the carriers
of two devices are interacting, then both devices should exhibit similar
nearness patterns, in terms of propinquity and social interaction. 

\begin{figure}
\subfloat[Usense5 to all neighbors.]{\includegraphics[scale=0.4]{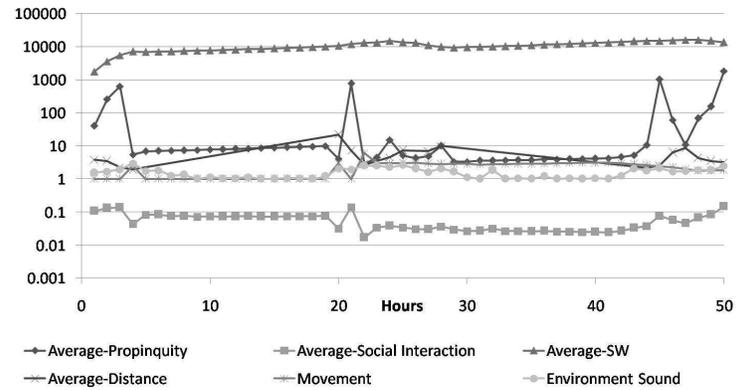}

}\\
\subfloat[USense3 to all neighbors.]{\includegraphics[scale=0.4]{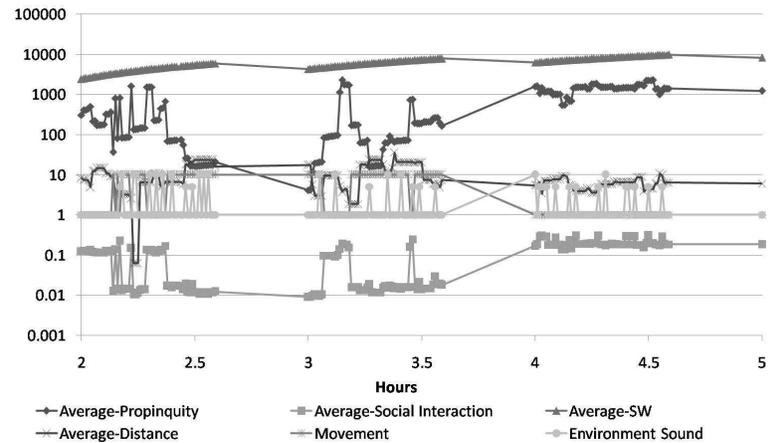}

}

\caption{Experiment III results, USense5 and USense3 perspective.\label{fig:Experiment-III}}
\end{figure}

Hence, considering again the raw traces extracted during Experiment
II, Figure \ref{fig:Experiment-III} covers results for a 3-hour observation
period, between hours 2 and 5 (24.11.2015, 4 p.m. and 7 p.m.), a period
which has been randomly selected from the specific periods where 2
devices, USense3 and USense5 interacted the most over the period of
50 hours of Experiment II. Figure \ref{fig:Experiment-III}a) holds
results of USense5 towards all of its neighbors, while Figure \ref{fig:Experiment-III}b)
holds results concerning USense3 towards all of its neighbors.

In all experiments performed under this scenario we have observed
symmetry in terms of social interaction. Similarly, for this particular
case, the captured social interaction pattern holds symmetry. However,
in what concerns propinquity, such property is now weaker. For instance,
between hours 2 and 2.5, propinquity exhibits more variability for
USense3 than for USense5. The reason for this concerns the fact that
propinquity is more significantly affected by distance fluctuations.
This is an aspect that we expected to further analyze in future research,
as our tool exhibited some limitations (delay) in computing the distance
between neighboring nodes. 

\section{Conclusions and Future Work}

\label{sec:Conclusions}

This paper discusses a tool that can assist in a non-intrusive way
a contextualization of nearness, based on two specific utility functions
that model social interaction and propinquity. Social interaction
modeling is relevant to understand whether or not nearness exists
at an instant in time; propinquity is relevant to assist in a longer-term
characterization of such nearness as well as in assisting to detect
abnormal patterns of behavior (e.g. isolation).

The paper presents a tool, NSense, that has been used under realistic
conditions to validate the proposed concepts of social interaction
and propinquity. We have validated such tool under realistic settings,
showing that the proposed functions are sound for short-time periods
(e.g., 7 hours) as well as for longer time periods (50 hours) and
relevant in a characterization of nearness. Results obtained corroborate
that a nearness context can be based on the proposed functions, as
they assist in understanding patterns of social interaction; resting
periods; abnormal patterns (e.g., social isolation). The proposed
functions are relevant in terms of local (end-user device) inference,
and robust over time. We believe that the results that they provide
can be improved by adequate individual training, an aspect that is
being already addressed. A second conclusion to draw from the experiments
carried out is that nodes interacting exhibit symmetric patterns of
social interaction. While for the case of propinquity, the pattern
symmetry involving nodes that interact regularly seem to be weaker
- an aspect that we believe may be improved by creating a method that
computes the direct distance via Wi-Fi quicker.

As ongoing work, we expect to explore this nearness context modeling
to, together with stronger psychological assessment and validation,
assist in promoting social interaction between familiar strangers.
A second line of work that we are developing concerns the fundamental
role that social interaction and propinquity may have as early indicators
of social anxiety.

\section{Acknowledgment}

Thanks are due to the COPELABS CitySense project. The paper acknowledges
support from the H2020 UMOBILE project, grant number 645124.

\section{References}

\bibliographystyle{ieeetr}
\bibliography{bib-or}

\end{document}